\documentclass[preprint,amsmath,amssymb,a4]{revtex4}    
\usepackage{epsfig}    
\usepackage{graphicx}    
\usepackage{rotating}    
\usepackage{amssymb}    
\usepackage{amsmath}    
\usepackage{subfigure}    
\usepackage{multirow}    
\def\beq{\begin{eqnarray}}    
\def\eeq{\end{eqnarray}}    
\def\be{\begin{equation}}    
\def\ee{\end{equation}}

\begin{document}    
    
\title{Balanced K-SAT and Biased random K-SAT on trees}    
    
\author{Sumedha$^{(1)}$, Supriya Krishnamurthy$^{(2,3)}$ and Sharmistha Sahoo$^{(1)}$}    
\affiliation{(1): National Institute of Science Education and Research, Institute of Physics Campus, Bhubaneswar, Orissa- 751 005, India\\(2): Department of Physics, Stockholm University, SE- 106 91, Stockholm, Sweden \\ (3): School of Computer Science and Communication, KTH, SE- 100 44 Stockholm, Sweden}   
   
\date{\today}   
\begin{abstract}   
We study and solve some variations of the    
random $K$-satisfiability problem - balanced $K$-SAT and biased random    
$K$-SAT -    
on a regular tree, using techniques we have developed earlier \cite{ss}.    
In both these problems, as well as variations of these that    
we have looked at, we find that the SAT-UNSAT    
transition obtained on the Bethe lattice matches the exact threshold    
for the same model on a random graph for $K=2$    
and is very close to the numerical value obtained for $K=3$.     
For higher $K$ it deviates from the numerical estimates of    
the solvability threshold on random graphs,    
but is very close to the dynamical $1$-RSB    
threshold as obtained from the first non-trivial fixed point of the    
survey propagation algorithm.

\end{abstract}   
\maketitle

Random $K$-Satisfiability is a random constraint satisfaction problem    
in which one tries to find a satisfying assignment for a randomly generated    
logical expression in conjugate normal form(CNF), which is an AND of    
$M$ clauses. Each clause consists of an OR of $K$ Boolean literals    
which are chosen randomly from a set of $N$ Boolean variables. As the    
constraint density ($\alpha = M/N$) increases, the number of satisfying    
assignments decreases. In the limit of $M \rightarrow \infty$ and     
$N \rightarrow \infty$, the system is known to have a sharp threshold     
in constraint density $\alpha_c$ below which the probability of finding     
satisfiable assignments approaches $1$ and above which it vanishes.    
\cite{msl,kirkpatrick}.     
      
The problem is originally  defined on a random graph, but    
because of the presence of loops, this is hard to     
solve exactly for arbitrary $K$.     
The location of a sharp threshold $\alpha_c$ is known rigorously      
for $K=2$ \cite{chvataletal}.     
For higher $K$ only bounds on upper and lower     
thresholds are proven\cite{achlioptas_bounds}.     
However using non-rigorous but powerful methods from statistical physics,      
namely the replica and cavity methods,  estimates for the     
threshold are obtained which seem to be very close     
to the values obtained numerically \cite{mezard-science,mmz,monasson}.     
    
The replica and cavity methods also predict that the solvability threshold     
is only one of many thresholds that exist in the problem     
as the number of constraints is increased. Before the solvability transition     
occurs, it is conjectured that     
the set of solutions (or satisfying assignments) first     
breaks up into a large number of well separated clusters     
at the clustering transition $\alpha_d$ \cite{mezard-science,biroli}.     
As the number of constraints further increases,     
it is argued that there is first a     
condensation transition \cite{lenkaetal_pnas} in which the number of clusters    
changes from from being exponentially numerous to     
sub-exponential and a freezing transition beyond which     
some variables take the same value in all the solutions of a     
given cluster \cite{semer,lk}.      
Though again, it is hard to prove rigorous results    
about the existence of these transitions on a random graph     
(however, in a recent result,\cite{coja-oghlan} proves rigorously  the existence of    
a clustering transition in some random constraint satisfaction problems) the cavity method is able to    
predict numbers very close to those observed numerically. In addition,    
it is conjectured \cite {mezard-montanari} that    
the threshold for clustering on a random graph is     
exactly equal to the reconstruction threshold on the corresponding tree    
(roughly defined, a tree graph is said to be reconstructable if the value     
that the root takes can be determined by the variable values at the leaves),    
supplying a further motivation for comparing results obtained thoretically     
on tree graphs with those obtained numerically for random graphs.

Recently we have studied the random $K$-SAT defined on a regular $d$-ary rooted     
tree \cite{ss}. By fixing the boundary we could calculate the moments of     
the number of solutions (averaged over all possible instances of the     
logical expression) exactly. Of more relevance to this paper, we     
also studied the probability (or fraction of instances) of     
having a satisfiable assignment as a function of tree depth. As expected,     
for a tree,  this probability may be written as a recursion     
relating the distribution at one level of the tree     
to the next level. We solved      
for the fixed point of these recursions to find that the behaviour of    
the probability  matches the behaviour of $K$-SAT on a random graph     
atleast qualitatively i.e, it shows a continuous transition for     
$K=2$ and a first order transition for $K \ge 3$.     
In addition we found that the value of $\alpha(K)$   
at which this transition takes place is very close to the value of the    
dynamical transition $\alpha_d(K)$ obtained for random graphs using the    
cavity method.    
    
In this paper we compute the same quantity (the fraction    
of satisfiable instances as a function of $d$)     
for some variants of the random     
$K$-SAT on a $d$-ary rooted tree and    
compare with the solvability thresholds obtained for the same problem on a    
regular random graph.     
We find that for $K=2$ the threshold obtained via the tree calculations     
matches the known exact value for a regular random graph. For higher $K$,     
we  have studied the problem numerically on regular random graphs     
and find that the threshold estimated on the tree is very close to the     
threshold obtained numerically for $K=3$, though    
deviating more and more as $K$ increases.     
Interestingly, as before \cite{ss}, the values obtained by our method     
are also very close to     
the value predicted for the clustering transition     
by dynamical $1$-RSB calculations on a random graph for these models    
(\cite{rm,zdeborova}).   
   
The two variants of $K$-SAT studied in this paper are    
Biased-random $K$-SAT and Balanced $K$-SAT.     
In biased random $K$-SAT each variable is negated with probability $p$ (for     
$p=1/2$ this reduces to the uniform random $K$-SAT). In Balanced $K$-SAT,     
each variable is constrained to occur negated and non-negated an equal     
number of times. We have also studied a generalisation of     
balanced $K$-SAT which we call $f$-balanced $K$-SAT, where a literal     
occurs $f$ times as one kind and $1-f$ as another.     

The plan of the paper is as follows:     
In Section \ref{section1} we explain the $d$-ary rooted tree on which we     
define various variants of random $K$-SAT. In Section \ref{section2} and     
\ref{section3} we study the biased random $K$-SAT and balanced $K$-SAT on     
the tree and compare the SAT-UNSAT threshold on the tree with the thresholds     
obtained on random graphs and regular random graphs. In Section     
\ref{section4} we explain the connection between our approach and the 
survey-propagation algorithm, which goes towards understanding why the numbers     
we get as an estimate of the solvability transition, also happen to be    
very close to the numbers obtained via dynamical 1-RSB calculations for the    
clustering transition.

\section{The model}    
\label{section1}    
\begin{figure}[tbp]    
\centering    
\includegraphics[scale=0.6]{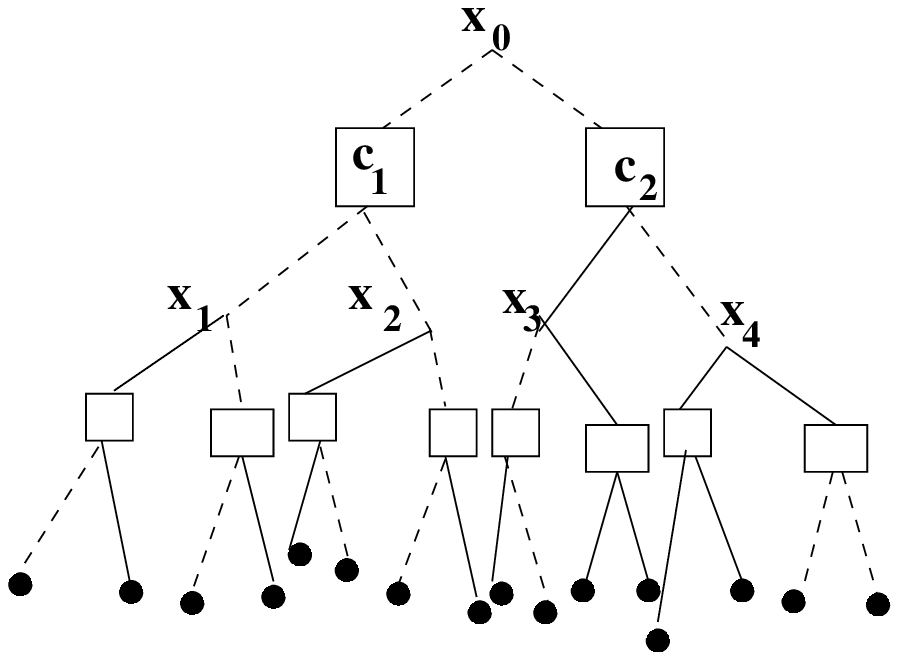}    
\caption{$3$-SAT on a rooted tree of depth $2$ and $d=2$.     
Only the clauses neighboring the root are labelled.    
Surface variables (or leaves) are depicted by a $\bullet$.     
Variable $x_0$  is at depth $2$, variables $x_1- x_4$ are at depth $1$ and the leaves are at depth $0$.     
Dashed/full lines between a variable and a clause indicate that     
it is negated/non-negated.}    
\label{fig:sat}    
\end{figure}    
    
We define the $K$-SAT problem on a tree as follows. Consider a     
regular $d$-ary tree $T$  in which every vertex has exactly $d$     
descendents.  The root of the tree $x_0$  has degree $d$ and its $d$ edges     
are connected to function nodes $\{ c_1,c_2...,c_d\}$. Each function     
node has degree $K$, and each of its $K-1$ descendents     
$\{x_i=x_1,x_2,....x_{k-1} \}$ is the  root of an independent tree     
(see Fig. \ref{fig:sat} ). Hence the root has a degree $d$ while all     
the other vertices on the tree (except the leaves which have a degree $=1$)     
have a degree $d+1$. Each vertex can take only two values: $-1$ or $1$.     
Each function node is associated independently with a clause     
$\phi(x_0,x_1,...,x_{k-1}) = \ell_0 \bigvee \ell_1\bigvee....\ell_{k-1}$.     
Here $\ell_i$ is one of the two literals $x_i$ or $\overline{x_i}$,        
depending on whether $x_i$ is joined to the function node by a     
dashed or a solid line (see Fig. ~\ref{fig:sat}).     
     
An assignment $\sigma$ of all the variables on the tree is a solution iff $\phi =1$     
for all the  clauses on the tree. One configuration of dashed and     
solid lines on the tree defines a realization $R$.     
    
We study the probability that a realization has no solution on this tree  
for a fixed boundary. A realization     
with no solution is one for which not a single assignment of the     
variables provides a solution. This  can happen if there is even a     
single variable on the graph, which, whether it takes the value $-1$ or     
$1$, causes atleast one clause to be unsatisfied. Such a variable then     
is a variable that can take $0$ values by our definition, and a     
realization that is not solvable has at least one variable of this     
type.      

On the tree graph, we can define the probabilities of a variable     
taking $0$, $1$ or $2$ values on the corresponding subtree.        
We define $p_i(0)$ as the conditional probability for a variable      
$x_i$ to cause a contradiction, in the subtree of which it is the root, given that all      
the other variables in the subtree can take at least $1$ value. We can     
then estimate the probability of a realization {\it having} a     
solution (or the fraction of realizations that have solutions) by  calculating     
the quantity $\Pi_{x_i} (1-p_i(0))$ where the product is over all the     
variables in the graph. The tree structure of the graph, gives us a     
way of calculating $p_i(0)$ through recursions.     
     
We define below     
some of the quantities in terms of which these recursions are written.     
We define $P_n(0)$ to be the probability that      
(or the fraction of realizations in which) a variable at depth $n$     
can neither take the value $-1$ nor take the value     
$1$, without causing a contradiction in its subtree.     
Here, by depth $n$, we mean a node that is $n$ levels away from the leaves    
(see Fig. ~\ref{fig:sat}).    
Note that because of the tree structure and because of the definition of the    
specific quantity we are looking at, all variables $x_i$ at depth $n$     
will have the same probability $P_n$. The probability that a     
variable at depth $n$, can take only one of the two values $-1$ or $1$ is     
defined to be $P_n(1)$ (the boundary nodes have $P_0(1)=1$, for example).   
Similarly  the probability that a variable at     
depth $n$ can take both values is $P_n(2)=1-P_n(0)-P_n(1)$.     
For the problems we look at, we are interested in the  
recursions for these quantities deep within the tree, as in \cite{ss}, so  
that we can get rid of boundary effects.

\section{Biased random K-SAT}    
\label{section2}    
We consider a model where each variable has a fixed degree $d+1$, and a     
variable appears as negated with probability p. For $p=1/2$ this corresponds     
to the uniform random K- SAT \cite{ss}. This model was also studied in     
\cite{rm} for $K=3$ on random graphs using the replica and cavity methods.    
    
\subsection{Biased random $K$-SAT on a tree}    
    
Let us first calculate $P_{n+1}(0)$ for variable $x_0$ (assuming it is at     
depth $n+1$), given these quantities for its descendents.     
Assume variable $x_0$ has a degree $d$ (by definition) and    
assume it is not negated on $d_1$ of these clauses. Variable $x_0$ will     
not be able to take the value $-1$ in the case when {\it at least} one     
of the $d_1$ clauses is {\it not} satisfied by the $K-1$ variables at the 
other end. In this case there will be at least one unsatisfied clause if     
$x_0$ takes the value $-1$. Similarly, if at least one of the $d-d_1$     
clauses which are satisfied by $x_0$, are also     
not satisfied by the $K-1$ variables at the other end, then $x_0$ cannot take     
the value $1$ either.    
    
It is easy to see that averaging over all realizations at     
depth $n+1$ implies averaging over all values of $d_1$, as well as averaging     
over all realizations  at depth $n$. It is important to note however     
tha realizations at depth $n+1$ are only built up from those     
realizations at depth $n$ that do have solutions. We define $Q_n$ as     
the conditional probability that a depth $n$ variable does not satisfy the     
clause above (to depth $n+1$), given that it has to be able to take at     
atleast one value (which satisfies the sub tree of which it is the root).     
The recursion for $P_{n+1}(0)$ is then:    
    
\begin{eqnarray}    
P_{n+1}(0) &=& \sum_{d_1=1}^{d_1=d-1} \left(\begin{array}{c} d \\ d_1     
\end{array}\right) p^{d_1} (1-p)^{d-d_1} (1-(1-Q_n^{K-1})^{d_1})     
(1-(1-Q_n^{K-1})^{d-d_1}) \nonumber \\    
&=& 1+(1-Q_n^{K-1})^d - (1-p Q_n^{k-1})^d - (1-(1-p) Q_n^{k-1})^d     
\end{eqnarray}    
     
Now we define $P_{n+1,-}(1)$ as the probability that a variable takes only     
one value and that value is $-1$ and $P_{n+1,+}(1)$ as the probability that     
the variable takes only one value and that value is $1$. Hence,     
the probability that a variable can take only one of the two possible     
values is $P_{n+1}(1) = P_{n+1,-}(1)+P_{n+1,+}(1)$. Recursions for     
these two quantities are as follows:    
\begin{eqnarray}    
P_{n+1,-}(1) &=& \sum_{d_1=0}^{d_1=d} \left(\begin{array}{c} d \\     
d_1 \end{array}\right) p^{d_1} (1-p)^{d-d_1} (1-(1-Q_n^{K-1})^{d_1}) ((1-Q_n^{K-1})^{d-d_1}) \nonumber\\    
&=& (1-(1-p) Q_n^{K-1})^d - (1-Q_n^{K-1})^d     
\end{eqnarray}

\begin{eqnarray}    
P_{n+1,+}(1) &=& \sum_{d_1=0}^{d_1=d} \left(\begin{array}{c} d \\ d_1 \end{array}\right) 
p^{d_1} (1-p)^{d-d_1} (1-(1-Q_n^{K-1})^{d-d_1}) ((1-Q_n^{K-1})^{d_1}) \nonumber\\    
&=& (1-p Q_n^{K-1})^d - (1-Q_n^{K-1})^d     
\end{eqnarray}    
    
Hence,    
\begin{equation}    
Q_{n+1} = \frac{p P_{n+1,+}(1) + (1-p) P_{n+1,-}(1)}{1-P_{n+1}(0)}    
\end{equation}    
This gives us the recursion   
    
\begin{equation}    
Q_{n+1}=\frac{p (1-p Q_n^{K-1})^d+(1-p) (1-(1-p) Q_n^{K-1})^d - 
(1-Q_n^{K-1})^d}{(1-p Q_n^{K-1})^d +(1-(1-p) Q_n^{K-1})^d -(1-Q_n^{K-1})^d}    
\label{qbias}    
\end{equation}    
These equations are a generalization of the recursions for $Q_{n+1}$     
obtained in \cite{ss} for $p=1/2$. From these equations the threshold at     
which the  fraction of realizations goes to zero exponentially with the     
depth of the tree may be extracted. This is the solvability threshold for     
these models \cite{kirkpatrick}. A fixed point analysis of Eq. \ref{qbias}     
predicts a continuous transition for $K=2$ and a first order     
transition for $K>2$ for all $0<p<1$ (We see no change in behaviour for any 
non-zero value of $p$ though, as reported in \cite{rm}). The value of $d$ at which     
the system undergoes a continuous transition for $K=2$ can be extracted     
by expanding to order $Q^2$ in Eq. \ref{qbias} at     
the fixed point. This gives, for $K=2$:    
\begin{equation}    
Q_c = \frac{2 (1-2 d p+2 d p^2)}{3 (d-1) d (p^2-p)}    
\label{k2p}    
\end{equation}    
which implies $d_c= \frac{1}{2 p (1-p)}$.    
    
\subsection{Comparison with results on random graph}    
\label{section1b}    
The calculations above should be compared with the value of the solvability     
threshold on a regular random graph. On general grounds, the value of $\alpha_c$    
corresponding to the fixed point value $d_c$ on the tree should be $(d_c+1)/K$    
\cite{baxter,gujrati}. Hence for $2$-SAT,    
$d_c=\frac{1}{2 p(1-p)}$ is equivalent to     
$\alpha_C =1/2 +1/(4p(1-p))$.     
   
A known earlier result \cite{cooper} provides us an opportunity to compare   
the above value of $\alpha$ with the exact value of    
the solvability threshold for $2$-SAT.    
Let $r_i$ represent the degree of $i^{th}$ variable on a random graph, and     
$r_{i,-}$ and $r_{i,+}$ be the degree of the corresponding literals.     
Hence $r_i = r_{i,-} + r_{i,+}$. For $2$-SAT defined  on a     
random graph with a  given literal distribution $R=\{r_{1,-},r_{1,+},....\}$,     
the location of the threshold can be derived using the following theorem     
by Cooper et al \cite{cooper}:    
    
{\it \bf Theorem:} Let $R$ be any degree sequence over $N$ variables,     
with $\Delta = N^{1/11}$, and let $F$ be a uniform random simple formula     
with a given degree sequence $R$, then for $0 < \epsilon <1$ and     
$N \rightarrow \infty$, if $D =\sum_i r_{i,-} r_{i,+} < (1-\epsilon)M$     
then P(F is satisfiable) $\rightarrow 1$ and if $D> (1+\epsilon)M$     
then P(F is satisfiable) $\rightarrow 0$.    
Here $M$ is the number of clauses.    
    
The theorem can be easily generalised to the case where we take D to be     
the average value of $\sum_i r_{i,-} r_{i,+}$     
in the case that the degrees of the variables are distributed according to     
a given probability distribution.     
    
For biased random $2$-SAT defined on a regular random graph,     
since the probability distribution of the literals is     
$p(r_{+}) = {r \choose r_{+}} p^{r_{+}} (1-p)^{r_{-}}$,   
we get $D=<r_{i,+} r_{i,-}>N=r(r-1)p(1-p) N$.     
Hence we get, $r_c =1+1/(2 p (1-p))$. Since $\alpha_c=r_c/2$ we get     
$\alpha_c = 1/(4 p (1-p)) +1/2$. This matches with the threshold     
obtained via the tree calculations (see Eq. \ref{k2p}).    
    
To compare the behaviour of biased random $2$-SAT on a random graph and on     
a regular random graph, we also calculated the threshold for biased random     
$2$-SAT on a random graph. In this case the degree of a variable is not fixed    
but is distributed according to a Poisson distribution. For this we get     
$D=<r_{i,+} r_{i,-}> N =N p (1-p) <r>^2$. The threshold then is given     
by the equation:    
    
\begin{equation}    
2 p (1-p) <r_c> =1    
\end{equation}    
    
Since $\alpha_c=<r_c>/2$, we get $\alpha_c=1/(4 p (1-p))$. As can be seen from 
the above result, the SAT-UNSAT transition threshold for Poisson distributed 
degree depends only on the average degree of the graph. Also, from the 
foregoing calculations, we see that for the same  $p$ the random $2$-SAT defined     
on a regular graph has a higher threshold. For example, for the most well     
studied case of $p=1/2$ the threshold value of $\alpha_c$ on random graph     
is $1$, while on regular random graph it is $3/2$.      
    
For $K>2$ there is no equivalent of the above theorem for  random graphs.     
Hence we  did numerical simulations for $K=3$ and $4$ for     
regular random graphs.     
While a lot of numercial work exists on $K$-SAT on random graphs    
\cite{msl,kirkpatrick}, random $K$-SAT on regular random graphs has not     
been studied much numerically. After generating $10^5$ random configurations     
of the logical expression, we count the number of solutions using     
the relsat algorithm\cite{bayardo}. Figs. 2 and 3 contain plots and     
finite size scaling data for $K=3$ and $4$ for $p=1/2$. We have compared     
the value of the threshold for regular random $3$-SAT on the tree and    
on random graphs in Table \ref{tab:a} for different values of $p$.     
Unlike $2$-SAT the values do not     
match exactly, but the tree calculations predict a threshold which is close to     
the threshold on a  regular random graph.    

\begin{table}     
\caption{Comparison of threshold for biased random regular $3$-SAT on a tree     
and on a regular random graph(RRG) for various values of $p$}     
\label{tab:a}     
	\centering     
		\begin{tabular}{|l|l|l|}     
		\hline	$p$ &  $tree$ & $RRG numerics $  \\    
		\hline   $0.5$ & $4.16$ & $4.36\pm0.03$ \\    
		\hline $0.45$ & $4.22$ & $4.45\pm0.05$  \\    
		\hline $0.4$ & $4.53$ & $4.75\pm0.05$   \\     
		\hline $0.35$ & $5.13$ & $5.30\pm0.05$  \\     
		\hline $0.3$ & $6.16$ & $6.28\pm0.05$ \\    
		\hline     
		\end{tabular}     
\end{table}     
    
For $p=1/2$ we have also compared the threshold obtained on the     
tree and from simulations of a  regular random  graph with that on a     
random graph (see Table \ref{tab:b}). As expected the difference between the     
model defined on a regular random graph and random graph goes down with     
increasing $K$. As we go to higher $K$, the mismatch between the threshold on     
the tree and a regular random graph increases. Interestingly, the     
value of the threshold obtained from tree calculations is very close to the     
value obtained via cavity method and 1-d RSB for the dynamical glass     
transition($\alpha_d$) \cite{mmz}. We will comment more on this in Section     
\ref{section4}.    
    
\begin{table}     
\caption{Comparison of threshold for random K SAT on a tree, regular     
random graph(RRG) and random graph(RG) for $p=1/2$. In the last column     
we have also reported values of $\alpha_d$ on a random graph. The starred     
values are exact values of the threshold as obtained using \cite{cooper}.}     
\label{tab:b}     
	\centering     
		\begin{tabular}{|l|l|l|l|l|}    
		\hline	$K$ & Tree &  RRG numerics &     
RG numerics(\cite{kirkpatrick})& $\alpha_d$ on RG(\cite{mmz})  \\    
		\hline   $2$ & $1.5$ & $1.5^*$ & $1^*$ & $1$ \\    
		\hline $3$ & $4.166$ & $4.36\pm0.03$ & $4.17\pm0.05$ & $3.93$   \\    
		\hline $4$ & $8.4$ & $9.86\pm0.03$ & $9.75\pm0.05$ & $8.3$\\     
		\hline     
		\end{tabular}     
\end{table}

\begin{figure}[ht]    
\begin{minipage}[b]{0.45\linewidth}    
\centering    
\includegraphics[width=\textwidth]{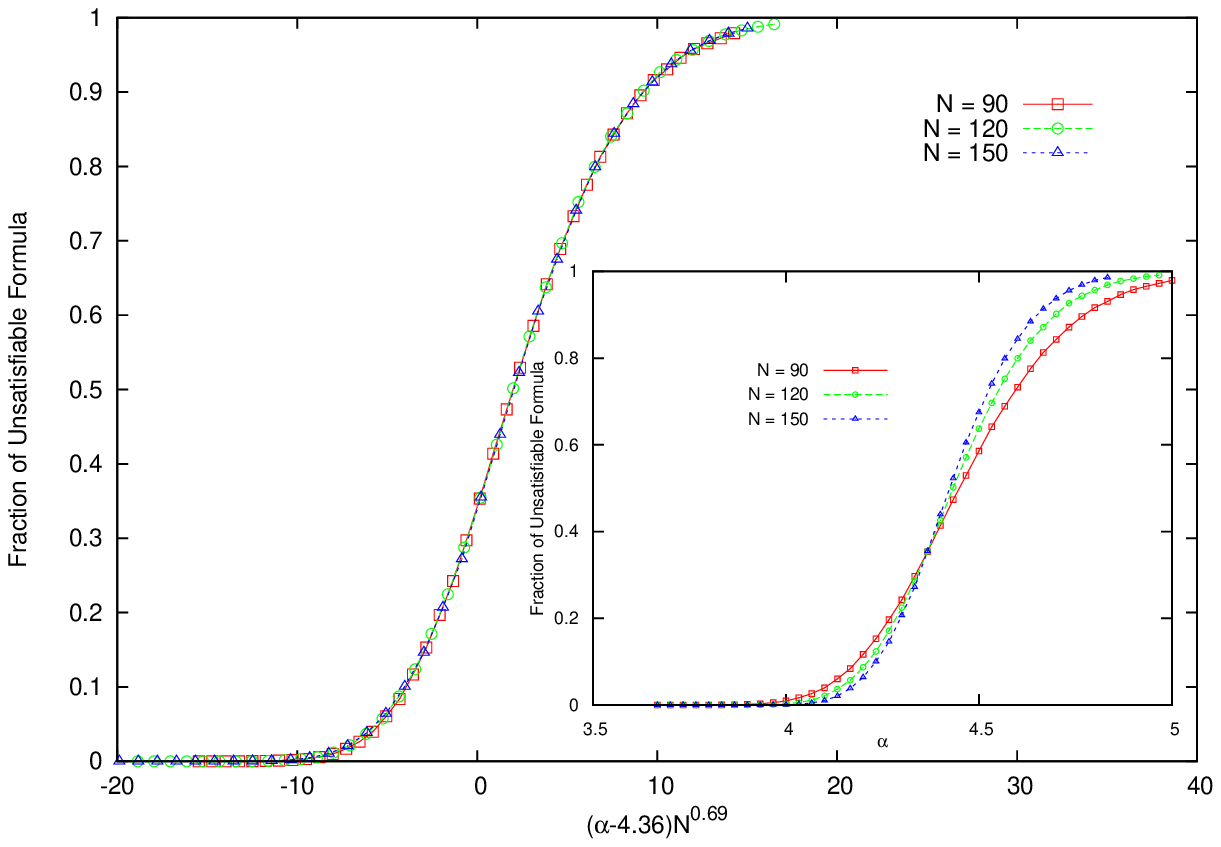}    
\caption{Scaled numerical data for $p=1/2$ for random $3$-SAT on a     
regular random  graph. The inset shows the unscaled value of the     
fraction of unsatisfied formulae as a function of $\alpha$.}    
\label{fig:figure1}    
\end{minipage}    
\hspace{0.5cm}    
\begin{minipage}[b]{0.45\linewidth}    
\centering    
\includegraphics[width=\textwidth]{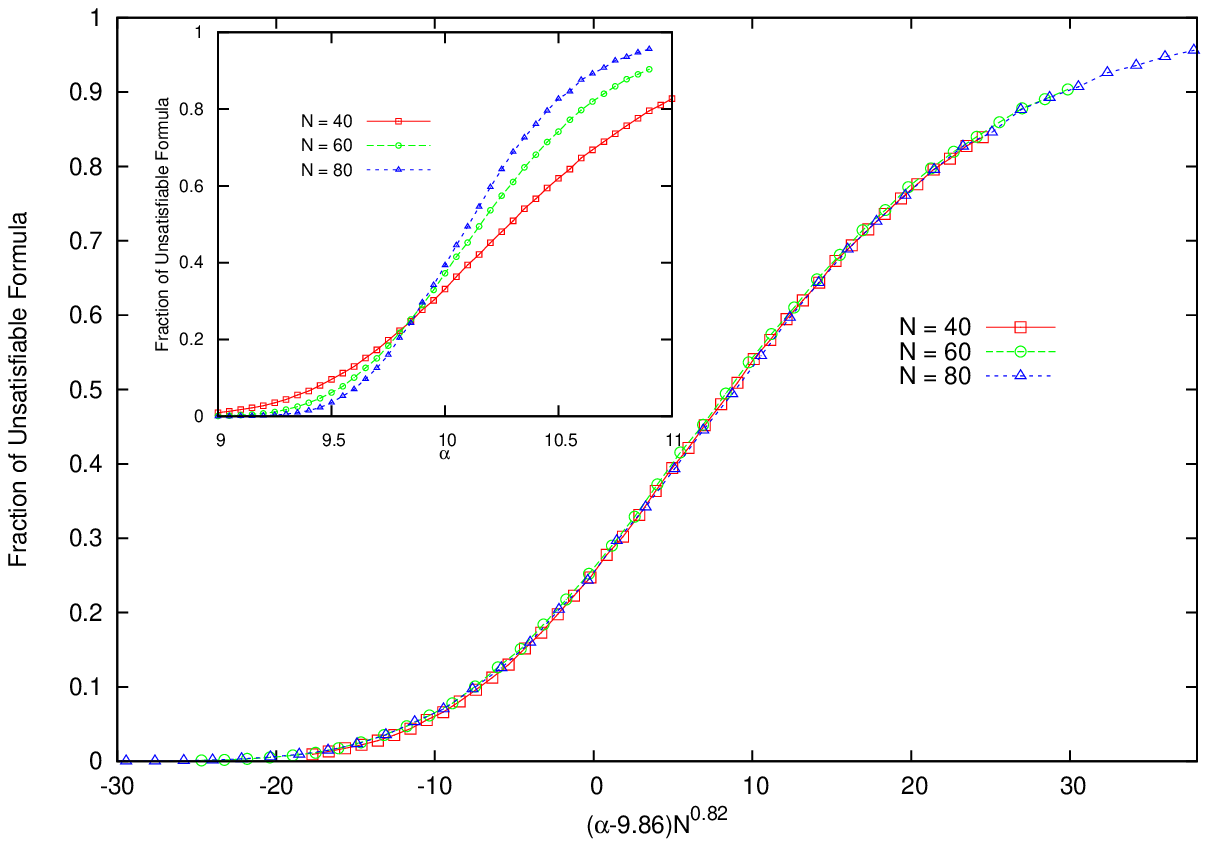}    
\caption{Scaled numerical data for $p=1/2$ for random $4$-SAT on a     
regular random graph. The inset shows the unscaled value of the     
fraction of unsatisfied formulas as a function of $\alpha$.}    
\label{fig:figure2}    
\end{minipage}    
\end{figure}    
    
\section{Balanced K-SAT}    
\label{section3}    
Balancing literals adds a dependency between variables, that complicates     
the problem. In Balanced $K$-SAT each literal is constrained to occur     
negated or non-negated exactly half the time. This model was shown to     
have higher complexity than random $K$-SAT \cite{boufkhad}. It is also a     
harder problem for standard SAT solvers as they depend on variable     
selection which exploits the difference in literal degrees.     
In the usually-studied version of the problem with $N$ nodes having an average    
degree $r$, the number of literals appearing with either sign is $Nr/2$.    
As mentioned earlier, apart from studying the above     
we also study a variant of the problem where the number of literals     
of one kind is $fNr$ while the number of the opposite kind is $(1-f)Nr$    
for any $0<f<1$. For  $f=1/2$, the problem is the usual one. For this case,     
bounds on the threshold have been derived in \cite{vish}  using the   
second moment method. For $K=3$ the problem has also been studied by   
Castellana and Zdeborov\'{a} \cite{zdeborova} using the cavity method.\\    
    
\subsection{Balanced regular $K$-SAT on a tree}    
    
Now besides, fixing the degree of variables to be $(d+1)$, we also fix     
the degree of the literals. Let a variable occur negated/non-negated exactly     
$(d+1)/2$ times. We define $q$ to be the integer value of $(d+1)/2$. We aim     
to write recursions for $Q_n$ as defined in the previous section.     
While the logic for writing these recursions is the same    
as before, the subtlety here is that, because of the balancing condition,     
whether a variable at depth  $n$ is negated or not in the clause     
connecting it to depth $n+1$ is not independent of whether it is     
negated or not in the other clauses it participates in.     
Nevertheless, our method is easily modified to deal with this    
situation. For ease of presentation we use the terms     
'downward' and 'upward' to    
denote a variable's connections to clauses at     
lower and higher depths respectively.    
Also, since the balancing condition crucially depends on whether     
$d$ is even or odd, we first do these two cases separately before    
presenting a general  formula valid for any value of $d$     
(including non integer values).    
    
\subsubsection{When d is odd}    
Since each variable occurs in $d+1$ clauses, the only realizations that     
are allowed are when it is negated and not negated in     
exactly $(d+1)/2 =q$ clauses. This leads to one literal occuring $q$ times     
and the other $q-1$ times amongst the downward clauses.     
The upward clause then contains the literal which appeared as a minority     
amongst the downward clauses.    
Since the two cases of whether the minority literal is a negation or     
a non-negation are entirely equivalent, it suffices to look     
at only one of these two cases. \\    

We need now to consider two situations separately - when the minority literal is    
 true or when the majority literal is true.    
In the former case, the variable is guaranteed to satisfy the upward clause    
while in the latter case, it is guaranteed to unsatisfy the upward clause.    
    
The equations for $P_n(0)$, $P_n(1)$ and $Q_n$ can be written as before.     
They are:    
    
\begin{equation}    
P_{n+1}(0)= (1-(1-Q_n^{K-1})^q) (1-(1-Q_n^{K-1})^{q-1})    
\end{equation}    
    
\begin{eqnarray}    
P_{n+1,-}(1) &=& (1-(1-Q_n^{k-1})^{q}) (1-Q_n^{k-1})^{q-1}\\    
P_{n+1,+}(1) &=&(1-(1-Q_n^{k-1})^{q-1}) (1-Q_n^{k-1})^{q}    
\end{eqnarray}    
    
and     
    
\begin{equation}    
P_{n+1}(1)= P_{n+1,-}(1) + P_{n+1,+}(1)    
\end{equation}    
    
Here $P_{n+1,-}(1)$ denotes the probability of the majority literal     
being true and $P_{n+1,+}(1)$ denotes the probability of the minority    
literal being true.    
The equation for $Q_n$ is then    
    
\begin{equation}    
Q_{n+1} = \left( \frac{P_{n+1,-}(1)}{1-P_{n+1}(0)} \right)    
\end{equation}    
    
A fixed point analysis of this equation exhibits     
a continuous transition for $K=2$     
and a discontinuous transition for $K \geq 3$. For $K=2$ the transition occurs    
at $d=1$. For $K=3$ and $d=7$ the fixed point equation has only one trivial     
solution ($Q=0$), while at $d=9$ it has three solutions,     
suggesting a first order  transition point in between these two values of $d$.    
    
\subsubsection{When d is even}    
In this case, its not possible to have exactly $(d+1)/2$ literals of     
one sign associated with every variable.     
Every variable has hence $d/2$ (or $q$)     
literals of one sign and $d/2+1$ (or $q+1$) literals of     
the opposite sign. Balancing is achieved by ensuring that    
for a graph of $N$ variables, exactly half the number of variables    
have, on average, $q$ literals of one sign while the other half have $q$    
literals of the opposite sign.    
    
As before, whether the minority number $q$ denotes negated or     
non-negated variables  is equivalent and we need only     
consider one of these cases.     
For a given sign of $q$ literals we need to consider again     
two distinct cases: All $q$ of the minority     
literals occur amongst the downward clauses or    
$q-1$ of the minority literals occur amongst the downward clauses and one in the    
upward clause. The former possibility occurs with probability $q/(2q+1)$    
and the latter with probability $(q+1)/(2q+1)$.     
The equation for $P_n(0)$ is now:    
\begin{equation}    
\label{eq:P0even}    
P_{n+1}(0)= (1-(1-Q_n^{K-1})^{q-1}) (1-(1-Q_n^{K-1})^{q+1})\frac{q}{2q+1}    
+ (1-(1-Q_n^{K-1})^{q}) (1-(1-Q_n^{K-1})^{q})\frac{q+1}{2q+1}    
\end{equation}    
The first term accounts for the case when all the majority literals occur    
amongst the downward clauses and     
the second term for the equivalent case when the    
minority variables all occur amongst the downward clauses.    
As before, for each of these two situations, the    
probability that the variable in question cannot take either value    
is  that at least one of the clauses this variable satisfies    
as well at least one of the clauses that this variable unsatisfies    
are also unsatisfied by the other variables which participate in them. \\    
    
Similarly, $Q_{n}$ is the probability (conditional on the node being able to take at least one value) that a node at level $n$    
takes the  one value that unsatisfies the upward clause.    
This happens when the node satisfies either the majority or minority     
literals which all occur amongst the downward clauses.    
    
This gives:    
\begin{eqnarray}    
\label{eq:Qeven}    
Q_{n+1} &=& \frac{q}{2q+1}\left( \frac{(1-(1-Q_n^{K-1})^{q+1}) (1-Q_n^{K-1})^{q-1}}{(1-P_{n+1}(0))}\right) \\    
&+& \frac{1+q}{2q+1}\left(\frac{(1-(1-Q_n^{K-1})^{q})(1-Q_n^{K-1})^{q}}{(1-P_{n+1}(0))} \right)     
\end{eqnarray}    
    
On solving for the fixed point, this equation indicates a continuous transition for $K=2$ between 
$d=0$ and $d=2$ and a first order transition for $K=3$ between $d=8$ and $d=10$.    
    
\subsubsection{For general d}    
Though the tree is defined for integer values of $d$, we can extend the    
above recursions to non-integer values. One way to do it is the folowing.    
For any arbitrary value of $d$, consider that a     
variable can occur negated in $q$ clauses a fraction $y$ of the times and    
in $q+1$ clauses a fraction $1-y$ of the times where $q$ is defined as before.     
The value $y=1/2$ corresponds to  even $d$ while $y=1$    
corresponds to odd $d$. So we have    
\be    
y q+(1-y)(q+1) = \frac{d+1}{2}    
\ee    
    
Note that the actual degree of the nodes is always $2q +1 $.     
So, for each variable,    
there is always one more of a literal of one sign over the other when     
$(d+1)/2$ is not an integer.    
The parameter $y$ ensures that on average the number of literals     
of either kind per node is always $(d+1)/2$, by fixing the fraction of nodes    
with one more negation over a  non-negation or vice versa.    
This procedure works for any $d$, including non-integral    
values since all that is needed is to fix $y$ accordingly from the     
above equation. \\    
    
The fixed point equation in this case     
for a general $y$ is exactly the    
same as Eq. \ref{eq:Qeven} for the case $y=1/2$.

Hence we can perform a  fixed point analysis of this equation for     
non-integer $d$. We get $d_c=1$ and hence $\alpha_c=(d_c+1)/2=1$ for     
$K=2$ and $d_c=8.65\pm 0.05$ for $K=3$ which gives     
$\alpha_c= (d_c+1)/3=3.23$.\\    
   
\subsubsection{f-balanced regular $K$- SAT}    
    
If instead of fixing the ratio of negated to non-negated     
to be $1/2$, we assume that it is some general fraction $f$, then     
again it is easy to write the fixed point recursion.    
For any general $f$, if $q$ is the integer value of $f(d+1)$, then we     
have to now consider two kinds of nodes:     
one for which the difference between the minority and    
majority literals is $d+1-q$ and the other for which the    
difference between the two is $d-q$.    
The value of $y$ fixes the fraction of these two kinds of nodes.  
The value $y=1-f$ corresponds to the case when $fd$ is an integer  
and the value $y=1$ corresponds to the case when $f(d+1)$ is an integer.  
For general $y$ we have,   
       
\be    
y q+(1-y)(q+1) = f(d+1)    
\ee    
    
The equations for $P_n(0)$ is now:    
    
\begin{eqnarray}    
P_{n+1}(0) &=& \left((1-F^{q-1}_{n,K}) (1-F^{d+1-q}_{n,K})\frac{q}{d+1}    
+ (1-F^{q}_{n,K}) (1-F^{d-q}_{n,K})\frac{d+1-q}{d+1}\right) y \nonumber\\    
&+&  \left((1-F^{q}_{n,K}) (1-F^{d-q}_{n,K})\frac{q+1}{d+1} +     
(1-F^{q+1}_{n,K}) (1-F^{d-q-1}_{n,K})\frac{d-q}{d+1} \right) (1-y) \nonumber    
\end{eqnarray}    
here we have defined $F_{n,K} =1-Q_n^{K-1}$ for ease of presentation.    
    
As before to get the fixed point equation for $Q_n$,      
we need only consider the cases when the literal that is satisfied    
occurs entirely amongst the downward clauses.    
    
\begin{eqnarray}    
Q_{n+1} &=& y \left( \frac{(d-q+1) (1-F^{q}_{n,K})F^{d-q}_{n,K}}{(d+1)     
(1-P_{n+1}(0))}+ \frac{q F^{q-1}_{n,K}(1-F^{d+1-q}_{n,K} }{(d+1)     
(1-P_{n+1}(0))}\right) \nonumber\\    
   &+& (1-y) \left( \frac{(d-q)(1-F^{q+1}_{n,K}) F^{d-q+1}_{n,K}}{(d+1)     
(1-P_{n+1}(0))}+\frac{(q+1)F^{q}_{n,K}(1-F^{d-q}_{n,K})}{(d+1)     
(1-P_{n+1}(0))}\right) \nonumber\\    
\label{f-bal}    
\end{eqnarray}    
  
\subsection{Comparison with random graph}    
    
Balancing the literals makes the problem more constrained. For a given     
distribution of degrees, $D=<r_+ r_->= f (1-f) <r^2>$ in the case when     
each literal is  chosen with one sign $f$ times and the other sign    
$(1-f)$ times.    
Applying the theorem described in Section \ref{section1b} results in the    
following threshold equation for f-balanced $2$-SAT:     
    
\begin{equation}    
2 f(1-f) <r^2> - <r>=0    
\label{bal2th}    
\end{equation}    
    
Hence in the balanced literal case the threshold is sensitive to the 
underlying distribution through the second moment. For $f=1/2$ we 
have the lowest threshold and the equation in that case is :    
    
\begin{equation}    
<r^2>-2<r>=0    
\end{equation}    
    
Interestingly this equation is exactly the same as the equation for  the    
percolation threshold on a random graph with a given degree distribution     
\cite{molloy1}. As argued by Molloy et al \cite{molloy2}, the SAT threshold     
cannot be lower than the percolation threshold. This implies that the most     
constrained $2$-SAT problem for a given degree/variable distribution is the     
one where the literals are exactly balanced ($f=1/2$).     
    
For balanced $2$-SAT on regular random graphs, $<r^2> =r^2$,     
and hence $\alpha_c = 1/(4 f (1-f)$. We have compared     
this with the threshold obtained from the fixed point analysis of     
Eq. \ref{f-bal} and it matches exactly. For example for $f=1/2$,      
we get $\alpha_c=1$.    
    
We have also considered balanced $2$-SAT with poisson distributed degree on a random graph. Unlike the     
regular random $2$-SAT, the threshold here for any arbitrary degree distribution     
depends on its second moment. For balanced $2$-SAT with Poisson distributed     
degree on a random-graph, $<r^2>= <r>+<r>^2$. Substituting in    
Eq. \ref{bal2th} gives $<r_c>= (\frac{1}{2 f (1-f)} -1)$ and hence     
$\alpha_c =1/(4 f(1-f)) -1/2$. This gives $\alpha_c=1/2$ for $f=1/2$.     
Note that this is also the percolation threshold for Erd\"{o}s-R\'{e}yni   
random graphs. This suggests that the model with balanced literals and   
a Poisson degree distribution on a random graph has the lowest     
SAT-UNSAT threshold among all possible models for $K=2$.     
    
For $K>2$, as in biased random K-SAT, the values obtained from the     
tree calculation seem to give a lower bound on the threshold value obtained     
numerically on regular random graphs (see Table \ref{tab:c}).    
We have also simulated     
balanced SAT on random graphs for $K=3$ and $4$ for $f=1/2$.     
Figs. 4 and 5 plot  the values of the fraction of unsatisfied formulas     
as a function of $\alpha$ for balanced $3$-SAT and $4$-SAT respectively,     
on a regular random graph. Figs. 6 and 7 show the same quantity for $3$-SAT     
and $4$-SAT on a random graph. Within numerical precision, the threshold     
on a random graph is lower than that on a regular random graph. As expected     
however, the difference deceases with increasing $K$.     
    
The problem of $K=3$ balanced SAT on regular and random graphs     
was studied using  belief propagation and survey propagation by     
Castellana and Zdeborov\'{a}\cite{zdeborova}. They find     
that on regular random graphs, survey propagation starts to     
converge towards a non-trivial fixed point for $r>9$. The corresponding value    
they find for random graphs is    
$\alpha>3.2$. Our calculations on a tree give a     
nontrivial fixed point at $d_c+1=9.65$, consistent with the results    
presented in \cite{zdeborova}. This gives us $\alpha_C =3.23$ on a tree.   
Again  this is very close to the $\alpha_d$ obtained in \cite{zdeborova}     
using survey propagation for balanced SAT on random graphs.    
    
\begin{table}     
\caption{Comparison of threshold for balanced $K$-SAT on tree, regular random     
graph(RRG) and random graph(RG)(See also Figs. 4,5,6 and 7). The starred     
values of the threshold obtained using \cite{cooper} in the table are exact.}    
\label{tab:c}     
	\centering     
		\begin{tabular}{|l|l|l|l|}    
		\hline	$K$ & Tree & RRG numerics & RG numerics  \\    
		\hline   $2$ & $1$ & $1^*$ & $1/2^*$ \\    
		\hline $3$ & $3.23$ & $3.5\pm0.02$ & $3.37\pm0.02$ \\    
		\hline $4$ & $7.163$ & $8.69\pm0.02$ & $8.65\pm0.02$ \\     
		\hline     
		\end{tabular}     
\end{table}

\begin{figure}[ht]    
\begin{minipage}[b]{0.45\linewidth}    
\centering    
\includegraphics[width=\textwidth]{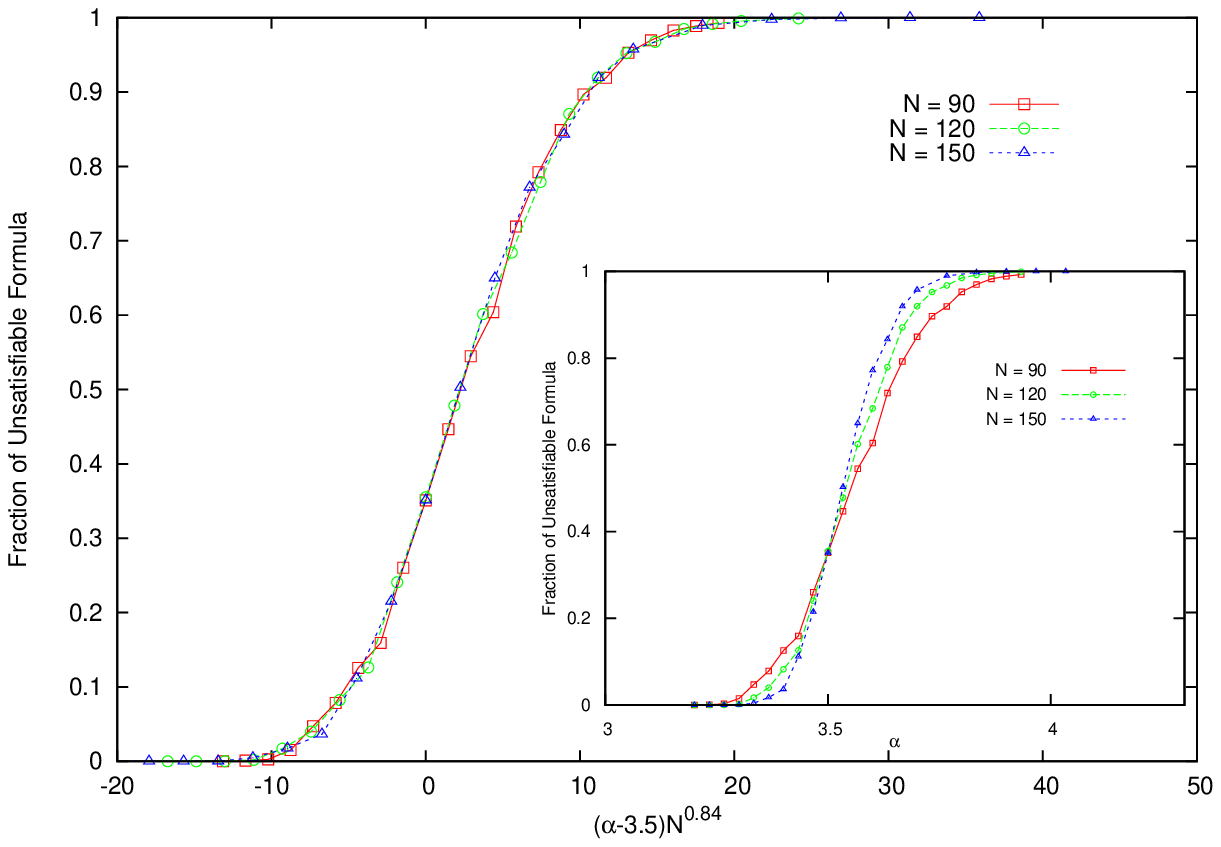}    
\caption{ Scaled numerical data for $f=1/2$ balanced $3$-SAT on a regular     
random graph. The inset shows the unscaled value of the fraction of unsatisfied     
formulae as a function of $\alpha$.}    
\label{fig:figure1}    
\end{minipage}    
\hspace{0.5cm}    
\begin{minipage}[b]{0.45\linewidth}    
\centering    
\includegraphics[width=\textwidth]{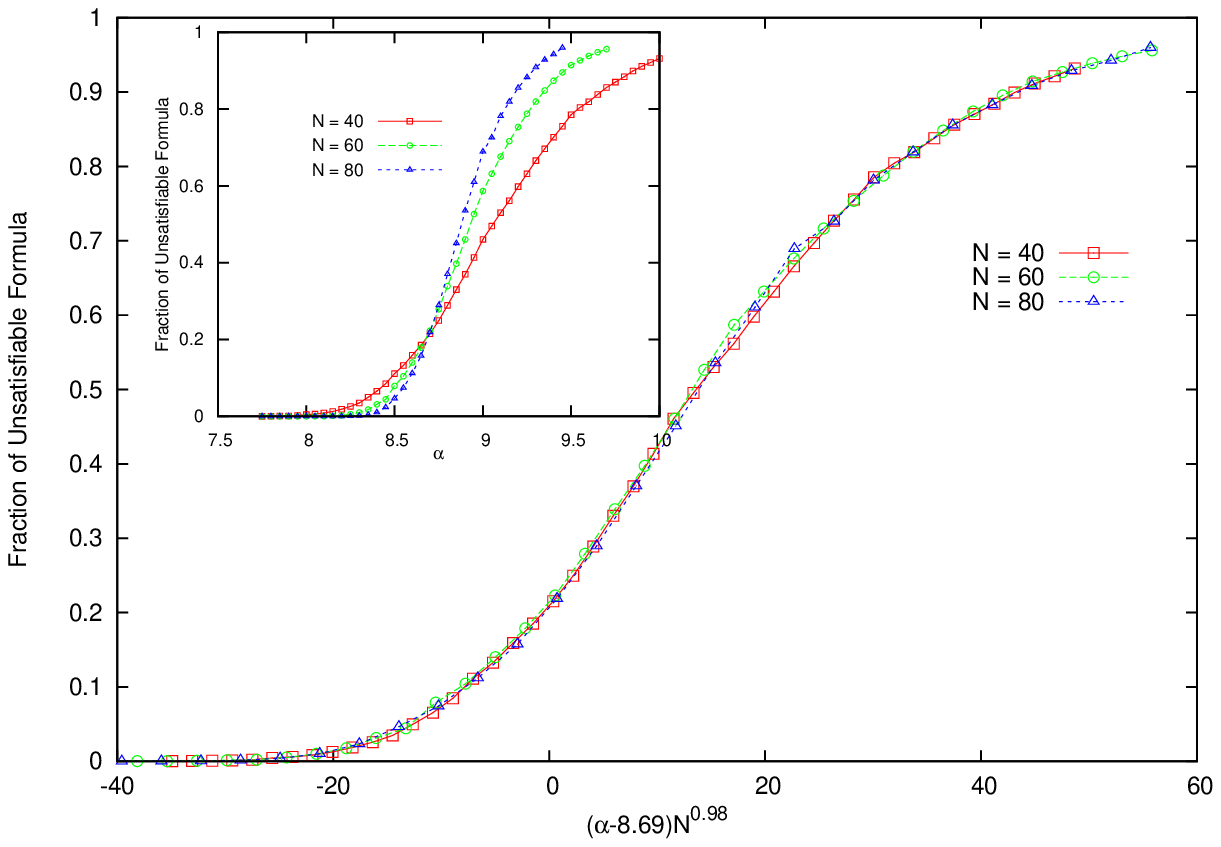}    
\caption{Scaled numerical data for $f=1/2$ balanced $4$-SAT on a regular     
random graph. The inset shows the unscaled value of the fraction of unsatisfied     
formulae as a function of $\alpha$.}    
\label{fig:figure2}    
\end{minipage}    
\end{figure}

\begin{figure}[ht]    
\begin{minipage}[b]{0.45\linewidth}    
\centering    
\includegraphics[width=\textwidth]{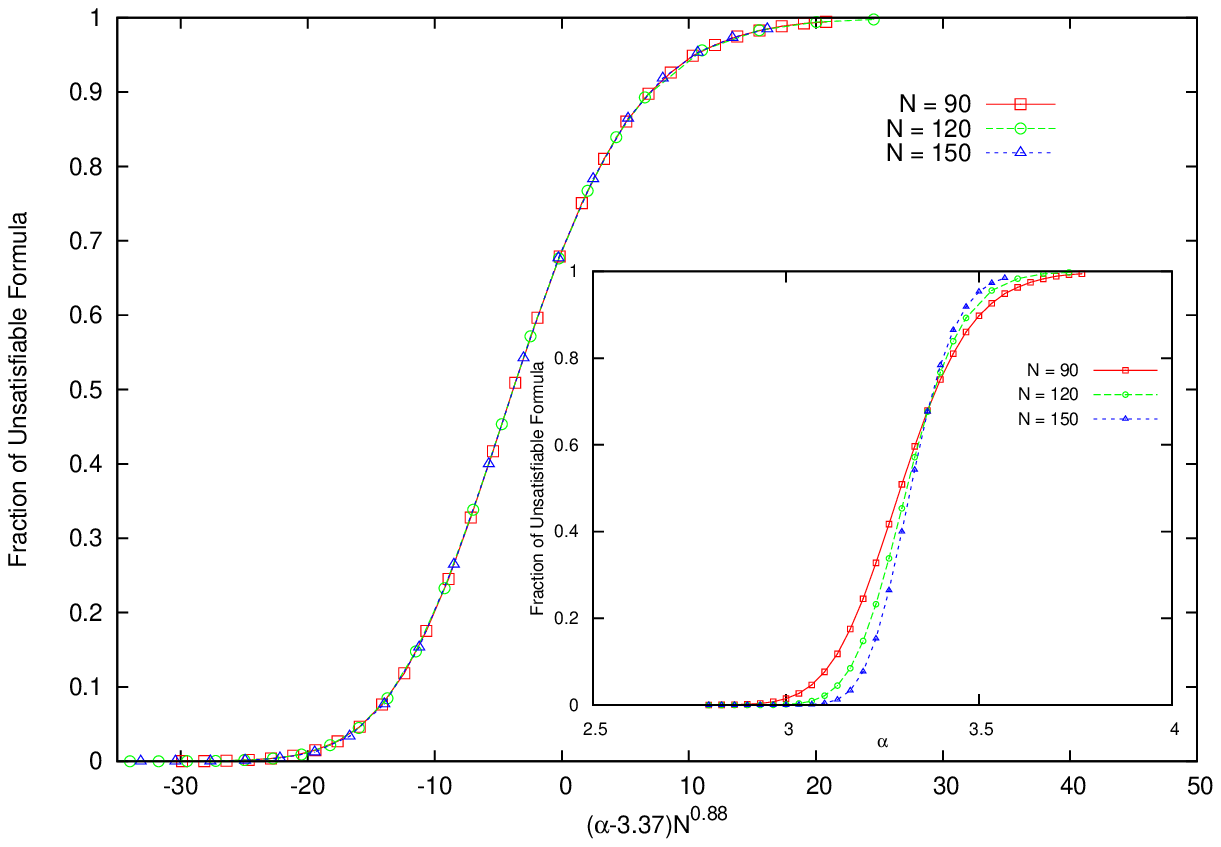}    
\caption{Scaled numerical data for $f=1/2$ balanced $3$-SAT on a random     
graph. The inset shows the unscaled value of the fraction of     
unsatisfied formulae as a     
function of $\alpha$.}    
\label{fig:figure1}    
\end{minipage}    
\hspace{0.5cm}    
\begin{minipage}[b]{0.45\linewidth}    
\centering    
\includegraphics[width=\textwidth]{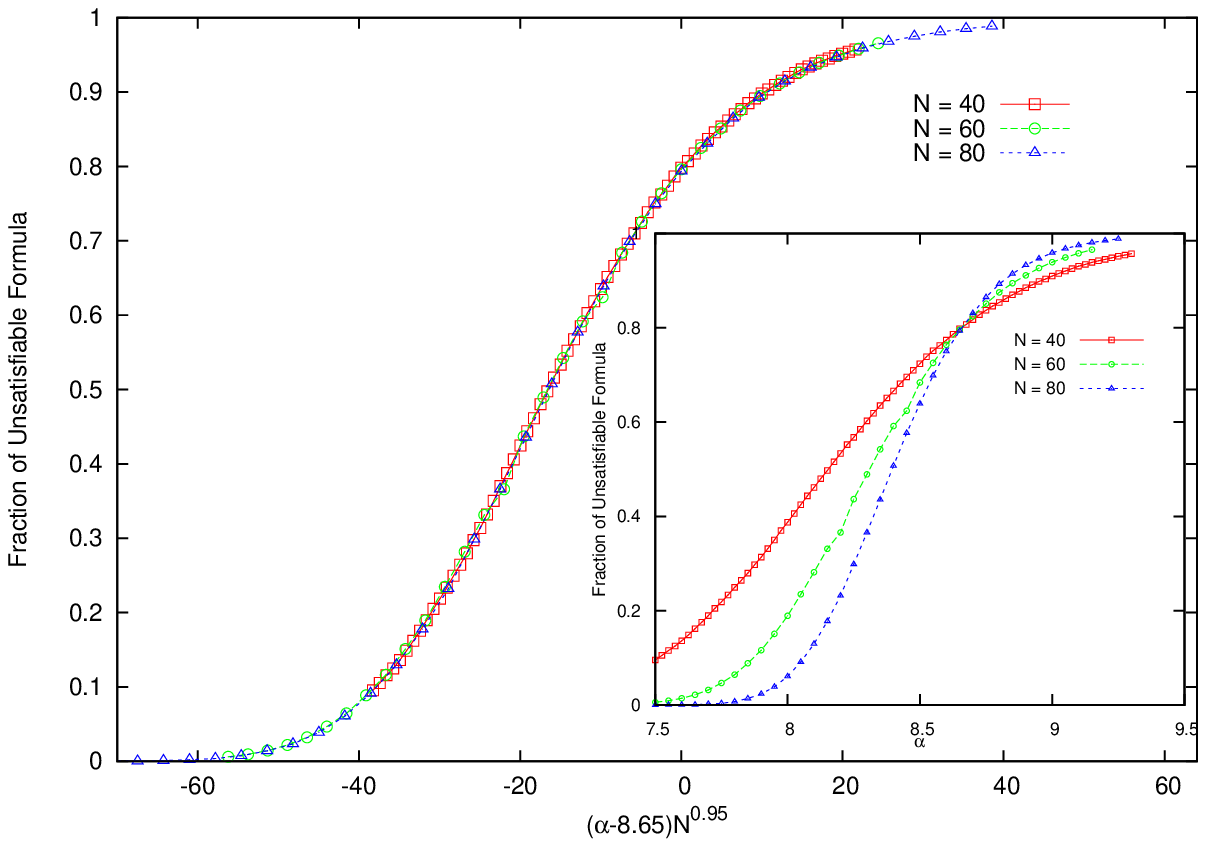}    
\caption{Scaled numerical data for $f=1/2$ balanced $4$-SAT on a random     
graph. The inset shows the unscaled value of the fraction of     
unsatisfied formulae    
as a function of $\alpha$.}    
\label{fig:figure2}    
\end{minipage}    
\end{figure}

\section{Connection with Survey Propagation and Reconstruction}    
\label{section4}    
  
The recursions we developed in \cite{ss} and in this paper are      
connected to the well-known problem of tree  reconstruction.      
The reconstruction problem, as originally defined, is a broadcast model     
on a tree, such that information is sent from the root to the     
leaves, across edges which act as noisy channels. The     
problem then is whether we can recover information     
about the root from a knowledge of the configuration of the leaves.    
Apart from its intrinsic interest, it is also of interest for   
$K$-SAT because it has been shown that the recursions     
developed in the reconstruction context are    
exactly the same as obtained by other means (such as the replica or     
cavity methods) for the dynamical glass transition on a random graph     
\cite{mezard-montanari} or the clustering transition     
for $K$ SAT (the value of $\alpha$ beyond which     
the solution-space is fragmented into different clusters).    
    
In terms of reconstruction, these fixed point     
recursions are developed for the    
unconditional probability distribution at the root of the tree    
to have a certain 'bias' ; namely, the fraction of      
boundary conditions (out of all boundary conditions      
that have a non-zero solution set), weigted by the total number of    
solutions these boundary conditions possess, that leads to the root     
taking the value $-1$ a certain number of times and the value $1$      
a certain number of times.     
    
The fixed point equations developed in \cite{ss} and in this paper      
have three differences in comparison with the one developed in     
\cite{mezard-montanari,bhatnagar-maneva}. We look at a reduced     
quantity - if the root can take two values      
(no matter what the bias), we lump it together to call it      
$P_n(2)$ for a level $n$.  The quantity of interest     
that we can now derive is a fixed point distribution for     
a single number, namely $Q_n \equiv P_n(1)/2(1-P_n(0))$.     
This makes our recursions similar in spirit to Survey propagation (SP)    
as we explain below.   
   
Secondly and more importantly, $P_n(1)$, $P_n(2)$ {\it etc}   
give the fraction      
of {\em realizations}, at the root, which have a     
non-zero solution set and {\em not}     
the fraction of boundaries.    
   
Thirdly,  unlike in  \cite{mezard-montanari,bhatnagar-maneva},    
where boundary conditions are weigted by the number of   
solutions they lead to, we do not weight the realizations   
by the number of solutions they have. Rather, to get $P_n(2)$ for example,   
we weight each realization equally in which the   
root can take both values  $1$ and $-1$.   
   
To see the similarities with earlier approaches better, let us   
now define the probability space    
over boundary conditions instead of realizations.   
We now derive a   
recursion for the  fraction of boundary conditions     
that fix the value unambiguously at the root (so that it is either   
$-1$ or $1$) at level $n$, given this quantity for level $n-1$. Only   
those boundary conditions that lead to solutions at level $n-1$ are   
permitted.  Note, as mentioned earlier, this is different from tree   
reconstruction in that now each boundary is weighted equally and not   
by the number of solutions it leads to.

These equations are the same as those derived earlier in      
\cite{ss}, since the constraints that lead to the recursions are   
the same (and are defined once we specify the model).   
The only difference is that, since we are working with a    
typical realization, the  extra average over all realizations    
is no longer allowed.  The equations are hence     
     
\begin{eqnarray}    
\label{eq:regKsat}    
P_{n+1}(0) &=& \left[ 1-\left(1- \left( \frac{P_n(1)}{2 (1-P_n(0))}\right)^{K-1}\right)^{d_1}\right] 
\left[ 1-\left(1- \left( \frac{P_n(1)}{2 (1-P_n(0))}\right)^{K-1}\right)^{d-d_1}\right] \nonumber \\     
P_{n+1}(1) &=& \left( 1 - \left(    
\frac{P_n(1)}{2 (1-P_n(0))} \right)^{K-1} \right)^{d-d_1} +    
\left( 1 - \left(    
\frac{P_n(1)}{2 (1-P_n(0))} \right)^{K-1} \right)^{d_1} - \nonumber \\   
& & 2 \left(1- \left(\frac {P_n(1)} {2 (1-P_n(0))} \right)^{K-1}\right)^d     
\end{eqnarray}    
where $d_1$ is a particular realization of negations at the root.     
The factor of $2$ in the expression      
$P_n(1)/(2 (1-P_n(0)))$ appears because if $P_n(1)$ is the total fraction      
of bc's  that determine the root (at level $n$) to be either $-1$ or $1$,      
then, because of symmetry, exactly half of these configurations will not     
satisfy the link to level $n+1$, no matter what this link is.       
    
If we replace $P_n(1)/(2 (1-P_n(0)))$ by $Q_n$ as before,      
and in addition replace $d_1$ by $~d/2$ to specify a {\em typical}     
realization, then we get the      
recursion     
    
\be    
\label{eq:modrec}    
Q_{n+1} = \frac{\left[1-Q_n^{K-1} \right]^{d/2} - \left[ 1-    
Q_n^{K-1}\right]^d}{{2 \left[1- Q_n^{K-1} \right]^{d/2} - \left[    
1- Q_n^{K-1} \right]^d}}     
\ee    
    
The fixed points for different $K$ obtained from the above      
equation are very close to the values obtained earlier in \cite{ss}.     
Infact, in the above form, we can also relate Eq. \ref{eq:modrec} to the    
form of the recursions derived in reference \cite{mmz} in their analysis    
of the SP algorithm. To see this, note that    
if we substitute $z= (1-Q^{K-1})^{d/2}$  in Eqn \ref{eq:modrec},     
we get the recursion    
    
\be    
\label{rec_sp}   
z = \left[1-\left( \frac{1-z}{2-z} \right)^{K-1} \right]^{d/2}    
\ee    
   
In the SP language, $Q^{K-1} $ is the same as the cavity bias survey and    
$z$ is the analog of the probability of receving no supporting    
(or impeding) warning.   
Eq. \ref{rec_sp} is exactly the recursion obtained in \cite{mmz},     
from the SP equations, when the probability    
distribution over the cavity bias surveys is replaced by a delta function,     
hence ignoring the differences in the values of these surveys    
between different variables $i$ or different realizations.    
    
In our case, we get the recursions quite simply and without any   
approximations, from the way we have set up the problem in terms of   
computing the fraction of solvable realizations. It is remarkable   
that these two different ways of thinking of the problem, one of   
which gives an estimate of the solvability transition and the other   
an estimate of the clustering transition, give the same recursions.   
   
In another interesting analogy, the recursions in Eq. \ref{eq:regKsat}   
are also exactly in the spirit of the 'naive reconstruction'   
algorithm mentioned by Semerjian \cite{semer} where  a connection   
is now made with the freezing transition.    
   
In conclusion our main contribution in this paper is that we have been able to get the 
exact SAT-UNSAT threshold for a number of models of random $K$-sat on a $d$-ary    
tree. This threshold matches exactly with the threshold on a random graph for     
$K=2$ and is very close to the numerical estimate of the threshold for     
$K=3$. In addition the numbers we get are equal to the numbers obtained for the    
dynamical glass transition for higher $K$ \cite{mmz}. The latter is a   
result of the fact that our equations though averaged over realizations,   
might equally well be thought of as an average over boundary conditions,   
and hence are connected to the analysis of the SP algorithm. However   
note that in our way of setting up the fixed point equations, we can     
directly make a connection with the solvability transition which, to   
our knowledge, has not been mentioned before in the context of a tree   
calculation. Usually the solvability transition is estimated via the    
complexity \cite{mmz}, which is defined as the number of constrained   
clusters in a typical instance of the problem. The complexity is   
calculated using the cavity method - it becomes non-zero at $\alpha_d$    
and reduces in value as $\alpha$ increases till it reaches $0$, which is   
the point conjectured to be the solvability transition. The values   
obtained by these means are very close to numerics for all values of   
$K$, unlike in our case. It would be interesting to understand whether   
any analog of the complexity can be formulated for the tree.   
    
Our fixed point equations are for a reduced or coarse-grained probability    
distribution function, but for this simplified quantity, we are able to   
write down an equation in closed form. It would be very   
interesting to  understand whether, in our formalism, the above is also   
possible for the full distribution, such as the distribution of the   
fraction of realizations (or boundary conditions) that the root takes   
the value $-1$ a fraction $\beta$ of the times and the value $1$ a   
fraction $1-\beta$ of the times. For either of these cases,   
weigting realizations by the number   
of solutions they possess is also an obvious generalization   
of the results presented here, which would be useful to investigate.  
  
Also, as mentioned here, variations of the same recursions seem to have   
connections to the clustering transition \cite{mmz}, the freezing   
transition \cite{semer} , and the solvability transition \cite{ss}.   
It would be useful to quantify this better, as a tree calculation  
being exact, would make it possibile to obtain precise bounds on these  
transitions.  
  
Our study of the different cases of balancing literals and degrees leads to the 
conclusion that balancing literals makes the problem harder while balancing the 
degree actually makes the problem easier. Hence the hardest    
problem, from the point of view of having the lowest SAT-UNSAT threshold    
is the case of balanced literals with a poisson degree distribution.    
In this case, for $K=2$ the solvability threshold is    
also the percolation threshold for Erd\"{o}s-R\'{e}yni random graphs,     
consistent with the conjecture that the satisfiability threshold on a graph    
cannot be lower than the percolation threshold on the same graph.

\end{document}